\documentclass[10pt]{article}

\usepackage{srcltx}

\usepackage{amsmath,amssymb,amsfonts}

\usepackage{epsfig}
\usepackage{color}

\usepackage{fullpage}
\usepackage{setspace}
\usepackage{flushend}

\usepackage{cite}
\usepackage{url}

\usepackage[short,12hr]{datetime}

       \usepackage{mathptmx}

\title{A DCT Approximation for Image Compression%
}

\author{%
R.~J.~Cintra%
\thanks{R. J. Cintra is 
with the 
Signal Processing Group,
Departamento de Estat\'istica,
Universidade Federal de Pernambuco, Recife, PE, Brazil.
Email: rjdsc@stat.ufpe.org}%
\and
F.~M.~Bayer%
\thanks{F.~M.~Bayer is with the 
Universidade Federal de Santa Maria, Brazil.
Email: bayer@ufsm.br}
}

\date{}

\begin{document}

\maketitle

\onehalfspacing

\begin{abstract}
An orthogonal approximation for the 8-point 
discrete cosine transform (DCT) is introduced.
The proposed transformation matrix contains only zeros and ones;
multiplications and bit-shift operations are absent.
Close spectral behavior relative to the DCT 
was adopted as design criterion.
The proposed algorithm is superior to the signed discrete cosine transform.
It could also outperform state-of-the-art 
algorithms in low and high
image compression scenarios,
exhibiting at the same time a comparable computational complexity.
\end{abstract}

\begin{center}
\end{center}

\section{Introduction}

The 8-point discrete cosine transform (DCT) is 
a key step in many image and video processing applications.
This particular blocklength is widely adopted in
several image and video coding standards, 
such as JPEG, MPEG-1, MPEG-2, H.261, and H.263~\cite{roma2007hybrid}.
This is mainly due to its good energy compaction properties,
which are closely related to 
the Karhunen-Lo\`eve 
transform~\cite{britanak2007discrete,effros2004suboptimality}.

During decades,
much has been done to devise fast algorithms
for the DCT.
This is illustrated in several prominent works including
\cite{suehiro1986fast,hou1087fast,arai1988fast,loeffler1991practical}.
In particular,
the DCT design
proposed by Arai~et~al~\cite{arai1988fast} became popular and
has been implemented in several different hardware architectures~\cite{dimitrov2004dctfree,arjuna2011algebraic}.
Nevertheless,
all these algorithms require several multiplication operations.
Past years have seen very few advances in the proposition of new
low-complexity algorithms for the \emph{exact} DCT calculation.
A possible exception is the arithmetic cosine transform,
whose mathematical background was recently proposed,
but much is yet to be developed in terms of 
practical implementation~\cite{cintra2010act}.

In this scenario,
signal processing community 
turned its focus to approximate
algorithms for the computation of the 8-point DCT.
While not computing the DCT exactly,
approximate methods can provide meaningful estimations
at low-complexity requirements.
Prominent techniques include
the
signed discrete cosine transform (SDCT)~\cite{haweel2001square},
the
Bouguezel-Ahmad-Swamy (BAS) series of 
algorithms~\cite{bouguezel2008multiplication,
bouguezel2008low,
bouguezel2009fast,
bouguezel2010novel,
bouguezel2011parametric},
and
the level 1 approximation by Lengwehasatit-Ortega~\cite{lengwehasatit2004scalable}.
All above mentioned techniques
possess extremely low arithmetic complexities.

In this context,
a new theoretical framework for DCT approximate transforms
was proposed by Cintra~\cite{cintra2011integer}.
The implied transformations are orthogonal
and are based on
polar decomposition methods~\cite{higham1986applications,cintra2011integer}.
The aim of this correspondence is to introduce
a new low-complexity DCT approximation
for image compression in conjunction with a quantization step.
After quantization,
the resulting approximate coefficients are expected to be close
to the ones furnished by the exact DCT.
We restrain our attention to matrices that are good DCT approximations.

\section{DCT round-off approximations}
\label{section.approximation}

The proposed approximation method modifies the
standard DCT matrix~$\mathbf{C}$ by means of the rounding-off operation.
Initially,
matrix~$\mathbf{C}$
is scaled by two and then submitted to
an element-wise round-off operation.
Let $[\cdot]$ denote the round-off operation as implemented
in Matlab programming environment~\cite{matlab2007}.
Thus,
the resulting matrix,
$\mathbf{C}_0 = [2 \cdot \mathbf{C}]$,
is shaped as follows:
\begin{align*}
\mathbf{C}_0
=
\left[
\begin{smallmatrix}
1 &\phantom{-}1 &\phantom{-}1 & \phantom{-}1 & \phantom{-}1 &\phantom{-}1 &\phantom{-}1 &\phantom{-}1 \\
1 &\phantom{-}1 &\phantom{-}1 & \phantom{-}0 & \phantom{-}0 &-1           &-1           &-1 \\
1 &\phantom{-}0 &\phantom{-}0 &-1            & -1           &\phantom{-}0 &\phantom{-}0 &\phantom{-}1 \\
1 &\phantom{-}0 &-1           &-1            & \phantom{-}1 &\phantom{-}1 &\phantom{-}0 &-1 \\
1 &-1           &-1           & \phantom{-}1 & \phantom{-}1 &-1           &-1           &\phantom{-}1 \\
1 &-1           &\phantom{-}0 & \phantom{-}1 & -1           &\phantom{-}0 &\phantom{-}1 &-1 \\
0 &-1           &\phantom{-}1 & \phantom{-}0 & \phantom{-}0 &\phantom{-}1 &-1           &\phantom{-}0 \\
0 &-1           &\phantom{-}1 &-1            & \phantom{-}1 &-1           &\phantom{-}1 &\phantom{-}0
\end{smallmatrix}
\right]
.
\end{align*}
Matrix $\mathbf{C}_0$ has some attractive computational properties:
(i) its constituent elements are $0$, $1$, or $-1$,
which is an indication of null multiplicative complexity;
(ii) as a transformation, it requires only additions, 
being bit-shift operations absent;
and
(iii) its scaled transpose can perform an approximate inversion,
making it a quasi-symmetrical tool (cf.~\cite{haweel2001square}).
In fact,
a coarse approximation for the DCT matrix is achieved by
$\hat{\mathbf{C}} = \frac{1}{2} \mathbf{C}_0$.
Notice that the presence of the scaling factor $1/2$ 
represents only bit-shifts.
In~\cite{cintra2011integer},
the scaling factor
that minimizes the Frobenius norm to the exact DCT matrix
was found to be
$0.3922$.

The good features of $\mathbf{C}_0$ could enable
such simple approximation matrix
$\hat{\mathbf{C}}$
to outperform the SDCT
in a wide range of practical compression ratios~\cite{bayer2010image}.
However, the suggested approximation has some drawbacks:
(i)~it lacks orthogonality,
since $\mathbf{C}_0^{-1} \neq \mathbf{C}_0^\top$,
where superscript ${}^\top$ denotes matrix transposition,
and
(ii)~its resulting approximation is poor when compared with some existing methods (e.g., the BAS algorithms).

This framework encourages a more comprehensive analysis of the discussed approximation.
Considering matrix polar decomposition theory~\cite{higham1986applications},
an adjustment matrix $\mathbf{S}$ that orthogonalizes
$\mathbf{C}_0$
is sought.
Indeed,
the referred orthogonalization matrix is given by
$\mathbf{S}
=
\sqrt{(\mathbf{C}_0 \cdot \mathbf{C}_0^\top)^{-1}}$,
where the matrix square root is taken in 
the principal sense~\cite[p.~20]{higham2008functions}.
This computation furnishes the diagonal matrix
$
\mathbf{S} = \mathrm{diag}
\left(
\frac{1}{2\sqrt{2}},
\frac{1}{\sqrt{6}},
\frac{1}{2},
\frac{1}{\sqrt{6}},
\frac{1}{2\sqrt{2}},
\frac{1}{\sqrt{6}},
\frac{1}{2},
\frac{1}{\sqrt{6}}
\right)
$.
Therefore,
the DCT matrix can be more adequately approximated
by the following proposed matrix:
$
\hat{\mathbf{C}}_\text{orth}
=
\mathbf{S}
\cdot
\mathbf{C}_0
$
.
Matrix~$\hat{\mathbf{C}}_\text{orth}$ possesses
useful properties:
(i)~it is orthogonal;
(ii)~it inherits the low computational complexity of $\mathbf{C}_0$;
and
(iii)~the orthogonalization matrix $\mathbf{S}$ is diagonal.

In terms of complexity assessment,
matrix $\mathbf{S}$ may not introduce an additional computational overhead.
For image compression,
the DCT operation is a pre-processing step for
a subsequent coefficient quantization procedure.
Therefore,
the scaling factors in the diagonal matrix $\mathbf{S}$ 
can be merged
into the quantization step.
This procedure is suggested and adopted in several 
works~\cite{bouguezel2008low,lengwehasatit2004scalable}.
As a consequence,
the computational complexity
of
$\hat{\mathbf{C}}_\text{orth}$ is ultimately confined to $\mathbf{C}_0$.

\begin{figure}
\centering
\input{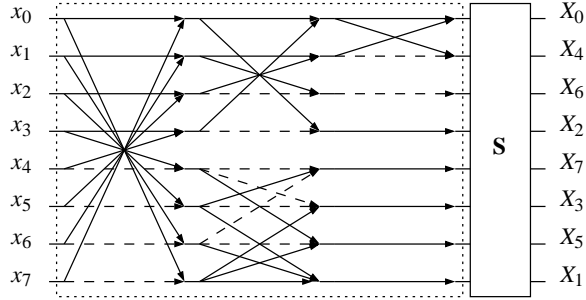}
\caption{Flow diagram of the proposed fast algorithm,
relating input data $x_n$, $n=0,1,\ldots,7$
to the corresponding approximate DCT coefficients $X_k$, $k=0,1,\ldots,7$.
Dotted box computes $\mathbf{C}_0$.
Dashed arrows represent multiplication by $-1$.}
\label{fig1}
\end{figure}

A fast algorithm for the transformation matrix $\mathbf{C}_0$
was devised and is depicted in Fig.~\ref{fig1}.
Arithmetic complexity assessment and comparisons are shown in Table~\ref{tab1}
in terms of addition, multiplication, and bit-shift counts.
The proposed algorithm is less complex than the SDCT 
and is only requires two additional operations
when compared to the best BAS algorithms.

\begin{table}
\centering
\caption{Arithmetic complexity analysis}
\label{tab1}
\begin{tabular}{lcccc}
\hline
Method & Add. & Mult. & Shifts & Total\\
\hline
Proposed Approximation
& 22 & 0 & 0 & 22 \\
BAS-2008 Algorithm~\cite{bouguezel2008low}
& 18 & 0 & 2 & 20 \\
BAS-2011 Algorithm~\cite{bouguezel2011parametric}
& 18 & 0 & 2 & 20 \\
SDCT~\cite{haweel2001square}
& 24 & 0 & 0 & 24 \\
BAS \cite{bouguezel2008multiplication}
& 21 & 0 & 0 & 21 \\
BAS \cite{bouguezel2009fast}
& 18 & 0 & 0 & 18 \\
BAS \cite{bouguezel2010novel}
& 24 & 0 & 4 & 28 \\
Level 1 Approximation~\cite{lengwehasatit2004scalable}
& 24 & 0 & 2 & 26 \\
\hline
\end{tabular}
\end{table}

For an initial comparison screening,
we separate 
(i)~the SDCT due to its very well documented 
literature~\cite{haweel2001square,britanak2007discrete};
(ii)~the BAS algorithms
introduced
in~\cite{bouguezel2008low} and \cite{bouguezel2011parametric}
(BAS-2008 and BAS-2011, respectively),
which are regarded the best BAS methods~\cite{bouguezel2009fast};
and
(iii)~the exact DCT.
The BAS-2011 algorithm was considered with its parameter set to $0.5$.
The approximation proposed 
in~\cite{lengwehasatit2004scalable}
and~\cite{bouguezel2010novel}
were not considered
because of their comparatively higher computational complexity
(Table~\ref{tab1}).

The proposed approximate matrix
is more closely
related to the DCT than the other approximations. 
In fact,
following the methodology suggested in \cite{haweel2001square}, 
the spectral structure, 
and energy compaction characteristics
could be assessed.
For such,
we understand each row of the transformation matrix as
coefficients of a FIR filter.
Thus,
the transfer function related to each
row of a given 
transformation
matrix $\mathbf{T}$
could be
calculated
according to
\begin{align*}
H_m(\omega;\mathbf{T})
= 
\sum_{n=0}^7 
t_{m,n}
\exp(-j n \omega),
\quad
m=0,1,\ldots,7,
\end{align*}
where 
$j=\sqrt{-1}$,
$\omega \in [0,\pi]$,
and 
$t_{m,n}$ is the $(m+1,n+1)$-th entry of $\mathbf{T}$.
A useful figure-of-merit is
the squared magnitude of the difference between
the transfer function of the DCT ($H_m(\omega; \mathbf{C})$) 
and 
of each considered approximation ($H_m(\omega; \mathbf{T})$).
This measure is clearly energy-related and
has the following mathematical expression:
\begin{align*}
D_m(\omega;\mathbf{T})
\triangleq
\left|
H_m(\omega;\mathbf{C})
-
H_m(\omega;\mathbf{T})
\right|^2
,
\quad
m=0,1,\ldots,7,
\end{align*}
where $\mathbf{T}$ is one of the selected approximate transforms.
For $m=0,4$,
the resulting transfer functions were the same for
DCT, SDCT, BAS-2008, BAS-2011,
and proposed approximation. 
Thus,
we restrict our comparisons to $m=1,2,3,5,6,7$.

\begin{figure}
\centering
\includegraphics[width=1.0\linewidth]{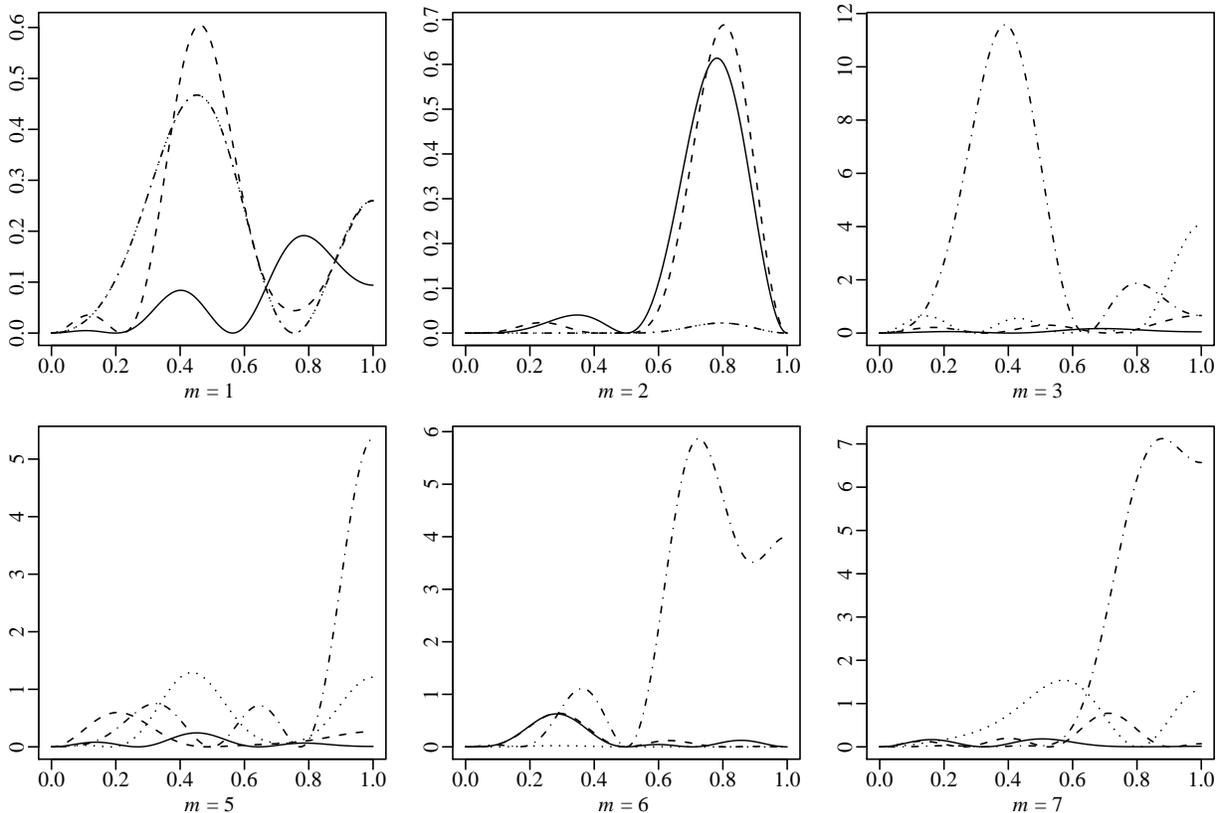}
\caption{Normalized plots of $D_m(\omega;\mathbf{T})$ 
for
the proposed algorithm (solid line), 
the SDCT (dashed line), 
the BAS-2008 algorithm (dotted line),
and
the BAS-2011 algorithm (dot-dashed line).}
\label{square_magnitude_error}
\end{figure}

Fig.~\ref{square_magnitude_error} 
shows strong similarities between
the spectral characteristics of the DCT and the proposed approximation.
The best results are presented when $m=1,3,5,7$. 
These spectrum similarity demonstrate 
the good energy related properties of the proposed algorithm.
Table~\ref{tab2} summarized the total error energy
departing from the actual DCT for each matrix row.
This quantity is given by
\begin{align*}
\epsilon_m(\mathbf{T})
=
\int_0^\pi
D_m(\omega;\mathbf{T})
\mathrm{d}\omega
\end{align*}
and was numerically evaluated.
On account of these results,
we remove the BAS-2011 method
from our subsequent analysis
because of its higher energy error.

\begin{table}
\centering
\caption{Error energy $\epsilon_m(\mathbf{T})$ 
for several approximate transforms}
\label{tab2}
\begin{tabular}{ccccc}
\hline
$m$ & BAS-2011 & BAS-2008 & SDCT & Proposed \\
\hline
1   &  0.59    &  0.59    & 0.59 & 0.21     \\
2   &  0.02    &  0.02    & 0.48 & 0.48     \\
3   & 10.64    &  1.93    & 0.59 & 0.21     \\
5   &  2.59    &  1.46    & 0.59 & 0.21    \\
6   &  6.28    &  0.02    & 0.48 & 0.48     \\
7   &  6.28    &  1.93    & 0.59 & 0.21     \\
\hline
Total
    &  26.40   &  5.93    & 3.32 & 1.79     \\
\hline
\end{tabular}
\end{table}

\section{Application to image compression and discussion}
\label{section.application}

The proposed approximation was assessed according to
the methodology
described in~\cite{haweel2001square} and
supported by~\cite{bouguezel2008low}.
A set of 45 $512\times 512$ 8-bit greyscale images obtained 
from a standard public image bank~\cite{uscsipi} was considered.
In this set,
the images employed in~\cite{bouguezel2008low} were included.

Each image was divided into $8\times8$ sub-blocks,
which were submitted to the two-dimensional (\mbox{2-D}) 
approximate transforms implied by
the proposed matrix~$\hat{\mathbf{C}}_\text{orth}$.
An 8$\times$8 image block $\mathbf{A}$ has its
\mbox{2-D} transform mathematically expressed by~\cite{suzuki2010integer}:
$\mathbf{T} \cdot \mathbf{A} \cdot \mathbf{T}^\top$,
where $\mathbf{T}$ is a considered transformation.
This computation furnished 64 approximate transform domain coefficients
for each sub-block.
According to the standard zigzag sequence~\cite{pao1998approximation},
only the $r$ initial coefficients were retained,
with the remaining ones set to zero.
We adopted $1 \leq r \leq 45$.
The inverse procedure was then applied to reconstruct the processed data
and image degradation is assessed.

\begin{figure}
\centering
\includegraphics[width=0.6\linewidth]{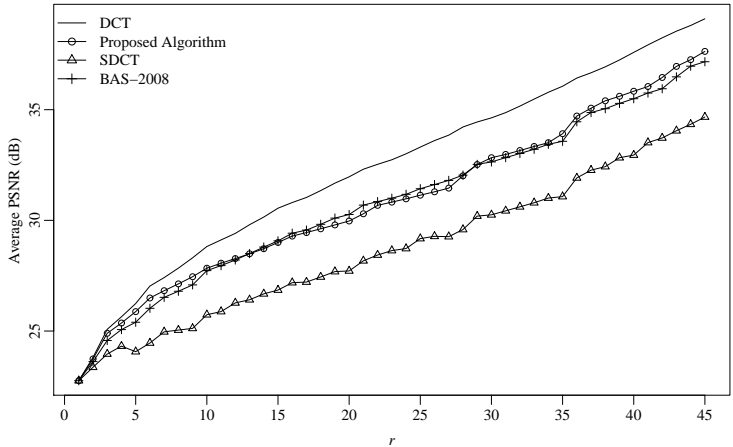}
\caption{Average PSNR for several compression ratios.}
\label{psnr}
\end{figure}

As suggested in~\cite{bouguezel2008low}, 
the peak signal-to-noise ratio (PSNR) was utilized as figure-of-merit.
However,
in contrast,
we are considering
the average PSNR from all images
instead of the
PSNR results obtained from
particular images.
Average calculations may furnish more robust results,
since the considered metric variance is expected to decrease as 
more images are analyzed~\cite{kay1993estimation}.
Fig.~\ref{psnr} shows that the proposed approximation 
$\hat{\mathbf{C}}_\text{orth}$
could indeed outperform the SDCT at any compression ratio. 
Moreover, 
it could also outperform 
the BAS-2008 algorithm~\cite{bouguezel2008low}
for high- and low-compression ratios.
In the mid-range compression ratios,
the performance was comparable.
This result could be achieved at the expense of 
only two additional arithmetic operations.

Additionally,
we also considered 
the universal quality index (UQI) 
and 
mean square error (MSE) 
as assessments tools.
The UQI is understood as 
a better method for image quality assessment~\cite{wang2002universal}
and 
the MSE is an error metrics commonly employed
when comparing image compression techniques.

\begin{figure}
\centering
\includegraphics[width=0.6\linewidth]{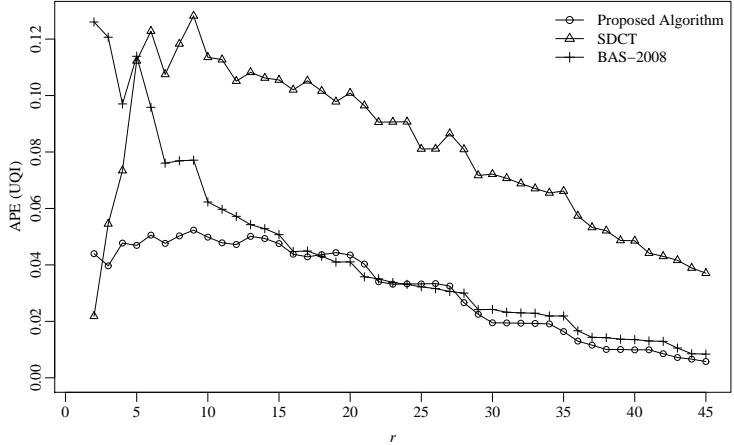}
\caption{Average UQI absolute percentage error relative to the DCT for several compression ratios.}
\label{uqi}
\end{figure}

\begin{figure}
\centering
\includegraphics[width=0.6\linewidth]{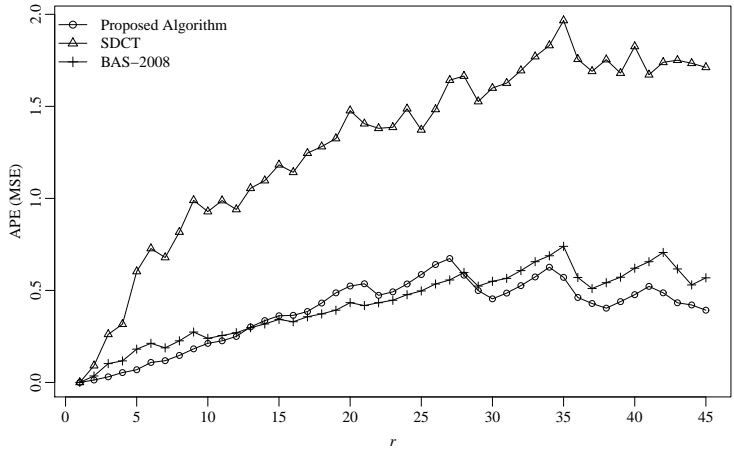}
\caption{Average MSE absolute percentage error relative to the DCT for several compression ratios.}
\label{mse}
\end{figure}

Fig.~\ref{uqi} and~\ref{mse} depict
the absolute percentage error (APE) relative to 
the exact DCT performance for the average UQI and average MSE,
respectively.
According to these metrics,
the proposed approximation 
lead to a better performance at almost all compression ratios. 
In particular, 
for high- and low-compression ratio applications the proposed approximation 
is clearly superior.

These results indicate that 
the proposed approximation is 
adequate for image compression, 
specifically for high-compression ratio applications. 
This scenario is found in low bit rate transmissions~\cite{bouguezel2008low}.
Additionally,
some applications operate with large amount of data
---
which demand fast, low-complexity algorithms
---
at
high compression ratios.
For instance,
models for face recognition and detection  
prescribe $r=15$ or 
less~\cite{sanderson2009multi,eickeler2000jpeg}.
This compression ratio is in one of the ranges
where
our proposed technique excels.
Additionally,
popular JPEG compression ratio are higher than 75\%,
which corresponds to $r\leq16$;
again a suitable requirement for the proposed approximation.

\begin{figure}
\centering
\includegraphics[width=1.0\linewidth]{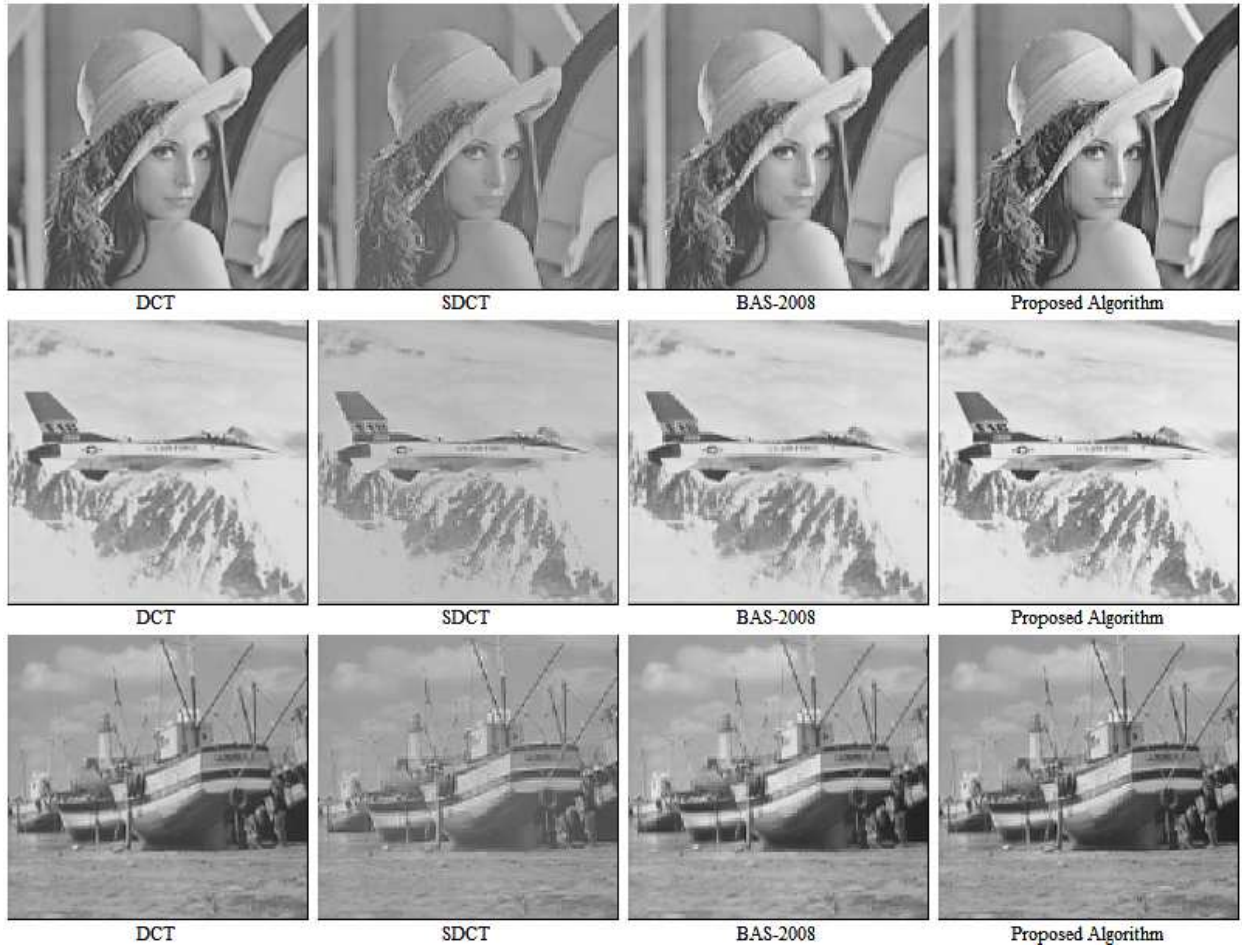}
\caption{Image compression using 
the DCT, 
the SDCT,
the BAS-2008 algorithm,
and
the proposed approximation 
for Lena, Airplane (F-16), and boat.512 images.}
\label{compress}
\end{figure}

Considering the above described compression methods,
we also provide a qualitative assessment of the
new approximation.
Using the 8$\times$8 block size
only 5~out of the 64~coefficients in each 8$\times$8 block 
were retained.
Thus,
after compression,
we derived the reconstructed images
according to 
the DCT, 
the SDCT, 
the BAS-2008 algorithm,
and
the proposed algorithm 
for three different standard images
(Lena, Airplane (F-16), boat.512) 
obtained from~\cite{uscsipi}. 
Fig.~\ref{compress}
shows the resulting images.
The resulting reconstructed images using the proposed method
are close to those obtained via the DCT. 
The superiority of the proposed algorithm over SDCT is clear. 
As expected,
the reconstructed images have better quality
and less blocking artifacts
when compared to the
BAS-2008 algorithm.

\section{Conclusion}
\label{section.conclusion}

This correspondence introduced an approximation algorithm
for the DCT computation
based on matrix polar decomposition.
The proposed method could outperform
the BAS-2008 method~\cite{bouguezel2008low}
in high- and low-compression ratios scenarios,
according to PSNR, UQI, and MSE measurements.
Moreover,
the proposed method possesses
constructive formulation based on the round-off function.
Therefore,
generalizations are more readily possible.
For example,
usual floor and ceiling functions can be considered 
instead of the round-off function.
This would furnish entirely new approximations.

Additionally,
the new approximate transform matrix has rows 
constructed from a different mathematical structure
when compared to the BAS series of approximations, 
for instance.
These rows can be considered in
the design of hybrid algorithms
which may take advantage of the best matrix rows from 
the existing algorithms
aiming at novel improved approximate transforms.
Finally,
the new approximation
offers the users another option for mathematical analysis
and
circuit implementation.

\section*{Acknowledgments}

This work was supported in part by 
Conselho Nacional de Desenvolvimento Cient\'ifico e Tecnol\'ogico, 
CNPq, Brazil, and FACEPE, Brazil.

\bibliographystyle{ieeetr}      %
\bibliography{rct2}   %

\end{document}